\documentclass[aps,prb,reprint,]{revtex4-2}

\usepackage{graphicx}
\usepackage{bm}
\usepackage{amsmath}
\usepackage{amssymb}
\usepackage{physics}
\usepackage{todonotes}
\usepackage{siunitx}
\usepackage{float}
\usepackage[hidelinks]{hyperref}
\usepackage[none]{hyphenat}
\usepackage{chemformula}

\newcommand{\kv}{\bm{k}}

\begin{document}
\title{Artificial oxide heterostructures with non-trivial topology}
\author{Pieter M. Gunnink}
\altaffiliation[Present address: ]{Institute for Theoretical Physics, Utrecht University, the Netherlands}
\email[Email at: ]{p.m.gunnink@uu.nl}
\affiliation{Faculty of Science and Technology and MESA+ Institute for Nanotechnology, University of Twente, 7500 AE Enschede, the Netherlands}

\author{Rosa Luca Bouwmeester}
\affiliation{Faculty of Science and Technology and MESA+ Institute for Nanotechnology, University of Twente, 7500 AE Enschede, the Netherlands}
\author{Alexander Brinkman}
\affiliation{Faculty of Science and Technology and MESA+ Institute for Nanotechnology, University of Twente, 7500 AE Enschede, the Netherlands}
\date{\today}

\begin{abstract}
In the quest for topological insulators with large band gaps, heterostructures with Rashba spin-orbit interactions come into play. Transition metal oxides with heavy ions are especially interesting in this respect. We discuss the design principles for stacking oxide Rashba layers. Assuming a single layer with a two-dimensional electron gas (2DEG) on both interfaces as a building block, a two-dimensional topological insulating phase is present when negative coupling between the 2DEGs exists. When stacking multiple building blocks, a two-dimensional or three-dimensional topological insulator is artificially created, depending on the \textit{intra-} and \textit{inter}layer coupling strengths and the number of building blocks. We show that the three-dimensional topological insulator is protected by reflection symmetry, and can therefore be classified as a topological crystalline insulator. In order to isolate the topological states from bulk states, the intralayer coupling term needs to be quadratic in momentum. It is described how such a quadratic coupling could potentially be realized by taking buckling within the layers into account. The buckling, thereby, brings the idea of stacked Rashba system very close to the alternative approach of realizing the buckled honeycomb lattice in [111]-oriented perovskite oxides.

\end{abstract}
\maketitle

\section{Introduction: topology by design}
After the experimental discovery of two-dimensional \cite{konig2DTI2007} (2D) and three-dimensional \cite{hsiehTopologicalDirac2008, chen3DTI2009} (3D) topological insulators in the class of chalcogenides many topological insulator predictions have been made for other materials classes, such as the Heusler compounds \cite{chadovTunableMulti2010}. All of these topological insulators have in common that the edges or surfaces have a linear Dirac dispersion and that the electron spin is locked to the momentum direction. Soon after, the notion of topology was generalized to include three-dimensional Dirac cones in the bulk: the Dirac and Weyl semimetals \cite{armitageWeylAndDirac2018}. The mechanism leading to the non-trivial topology is the spin-orbit interaction, hence the abundance of heavy atoms such as Bi or Hg in these topological materials.

Despite many advances, there is still a strong need for topological insulators with larger band gaps. This especially holds for two-dimensional topological materials with one-dimensional (1D) edge states, where band gaps are small \cite{du2008}. Both for the non-magnetic quantum spin Hall insulators (QSHI) \cite{konig2DTI2007} as well as the magnetic quantum anomalous Hall effect (QAHE) \cite{changExperimentalObservationQuantum2013,checkelskyTrajectoryAnomalousHall2014} interesting applications lay ahead once the effects would become robust at higher temperature. Therefore, alternative ideas to realize topology are being investigated, such as the use of electron-electron correlations or via the design of heterostructures. The material class of transition metal oxides is expected to be rich in effects when combining the two.

\begin{figure}[!b]
	\centering
	\includegraphics{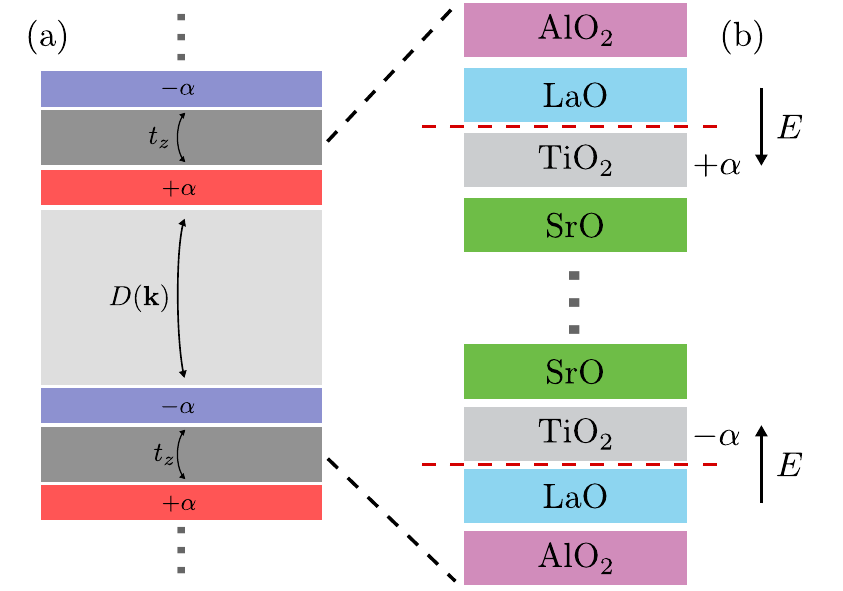}
	\caption{A schematic of an artificial topological insulator created by stacking oxides. (a) Shows the larger structure, where $-\alpha$ (purple) and $+\alpha$ (red) indicate 2DEGs with opposite Rashba splitting, $D$(\textbf{k}) the \textit{intra}layer coupling and $t_z$ the \textit{inter}layer coupling. In the example of the \ch{LaAlO3}/\ch{SrTiO3}/\ch{LaAlO3} building block, the light gray represents the \ch{SrTiO3} and the dark gray the \ch{LaAlO3}. (b) The termination switch is illustrated for the \ch{LaAlO3}/\ch{SrTiO3}/\ch{LaAlO3} heterostructure. The inversion symmetry is broken in the opposite way for the top and the bottom interface, which gives an opposite Rashba splitting. On the right the direction of the electric field is shown. The red dashed lines represent the 2DEGs. \label{fig:termination}}
\end{figure}

In recent years, much interest has been generated in topological crystalline insulators (TCIs) \cite{fuTopologicalCrystallineInsulators2011}. These are materials which exhibit topological features, but the topology is not protected by one of the three non-spatial general symmetries: chiral, particle-hole or time-reversal symmetry. Instead, the crystalline symmetry in combination with the aforementioned non-spatial symmetries introduces the topology and associated surface states. One would naively expect the surface states of these crystalline topological insulators to be immediately broken in the presence of a defect, since every crystallographic defect breaks the crystalline symmetry locally. Numerical simulations have, however, shown that the surface states of these TCIs are quite persistent \cite{mongQuantumTransportTwoParameter2012}. It turns out that if one averages over all disorders, the crystalline symmetry is restored, which allows for robust surface states \cite{andoTopologicalCrystallineInsulators2015}.

It is thus possible to design a topological insulator with a completely new toolbox, having a wide choice of materials and crystal structures. The first example of such a crystalline topological insulator was \ch{SnTe}, where only interfaces symmetric with respect to the \{110\} mirror plane have protected gapless states \cite{hsiehTopologicalCrystallineInsulators2012} There has been extensive work in designing topological insulators using these crystalline symmetries, for example by extending the Altland-Zirnbauer tenfold symmetry classes \cite{altlandNonstandardSymmetryClasses1997} with various crystalline symmetries \cite{shiozakiTopologyCrystallineInsulators2014}.

In this work, we suggest a topological crystalline insulator, expanding on the work previously done by Das and Balatsky \cite{dasEngineeringThreedimensionalTopological2013}, and we show how this model can be implemented using complex oxides. It is proposed to stack layers that have a strong Rashba spin-orbit splitting at their surfaces. We will explore whether this topological crystalline insulator can be realized in oxides by studying the example of a \ch{SrTiO3} layer embedded in \ch{LaAlO3}. First, we consider a single \ch{LaAlO3}/\ch{SrTiO3}/\ch{LaAlO3} heterostructure as building block, in which 2D topology is observed when the \textit{intra}layer coupling is negative. Subsequently, we stack multiple of these building blocks and observe that 2D and 3D topological insulators can be designed depending on the \textit{intra}- and \textit{inter}layer coupling strengths and number of building blocks. We will discuss the model and the symmetries leading to topology. Finally, we show how [111]-oriented oxide bilayers can also be used to realize a topological crystalline insulator.

\section{Stacking Rashba systems}
\label{sec:rashba-bilayers}
Our starting point for the design of a topological crystalline insulator is the model as introduced by Das and Balatsky \cite{dasEngineeringThreedimensionalTopological2013}. The goal of this model is to design a 3D topological insulator in a heterostructure, with the following characteristics: an insulating bulk, conducting surfaces and a Dirac-like dispersion with spin-momentum locking at the surface.

The basic building block for the heterostructure is a layer with a two-dimensional electron gas (2DEG) with Rashba-type spin-orbit splitting at both its interfaces. We choose the interfaces such that the Rashba splitting has an opposite sign for the top and bottom interface, which can be achieved by switching the inversion symmetry breaking between the interfaces \cite{petersenSimpleTightbindingModel2000}.
Perovskite oxides are particularly suited for this, since it is possible to control the termination layer atomically.~By switching the termination, we can reverse the inversion symmetry breaking, which switches the Rashba splitting. Throughout this work we will consider the example of a \ch{LaAlO3}/\ch{SrTiO3}/\ch{LaAlO3} heterostructure, as depicted in Fig.~\ref{fig:termination}. 

The \ch{LaAlO3}/\ch{SrTiO3} system is well known to have a 2DEG with Rashba splitting at its interface. The reason for the formation of a 2DEG in \ch{SrTiO3} when interfaced with \ch{LaAlO3}, is the polar catastrophe: half an electron moves from the \ch{LaAlO3} to the \ch{SrTiO3}, which dopes the \ch{SrTiO3} system and applies a strong electric field over the interface. This doping combined with the inversion breaking of the electric field induces a 2DEG with Rashba splitting \cite{caviglia2010}. Two 2DEGs can form within the same \ch{SrTiO3} layer, one at the top interface with \ch{LaAlO3} and one at the bottom interface with \ch{LaAlO3}. We will discuss the dynamics of the formation of two 2DEGs within this system in more detail in Sec.~\ref{sec:lao-sto-lao-poisson}.

During the fabrication of the heterostructure, an interface layer of \ch{LaTiO3} can be grown in order to switch the termination. Effectively, it is as if one grows half a unit cell in between one of the \ch{SrTiO3}/\ch{LaAlO3} interfaces. In this structure, the Rashba splitting of the top interface is opposite to that of the bottom interface, which is denoted in Fig.~\ref{fig:termination} by $\pm~\alpha$, because inversion symmetry is reversed for one of the interfaces. This is seen most easily by determining the electric field over the interface, which switches depending on the interface termination. A single \ch{LaAlO3}/\ch{SrTiO3}/\ch{LaAlO3} heterostructure, as shown Fig.~\ref{fig:termination}(b), is the building block for our larger heterostructure, in which we repeat a number of building blocks. We choose the \ch{LaAlO3} to be thin enough, such that the top interface of one building block interacts with bottom interface of the subsequent building block. 

One 2DEG is described by the Rashba Hamiltonian \cite{bychkovProperties2DElectron1984}
\begin{equation}
h_R^\pm = k^2/2m^* \pm \alpha_R ( \kv \times \bm{\sigma}),
\label{eq:ham-bilayer}
\end{equation}
where $\kv=(k_x, k_y)$, $k=|\kv|$, $m^*$ is the effective mass, $\alpha_R$ is the Rashba-coupling parameter and $\bm{\sigma}$ is the Pauli vector. 

\begin{figure}
	\includegraphics{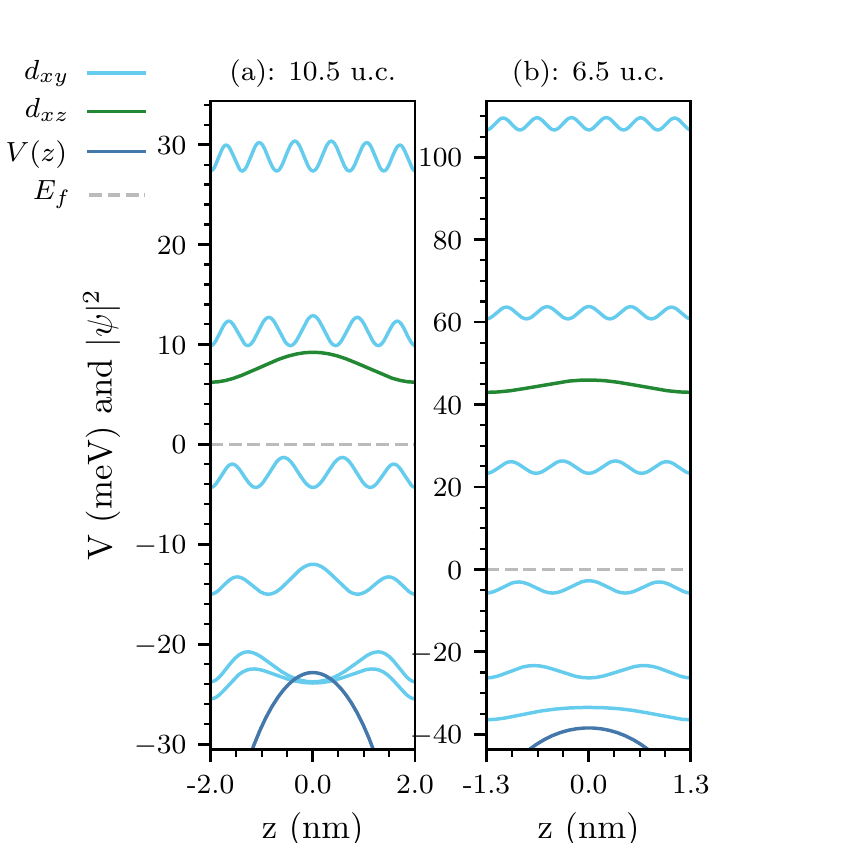}
	\caption{The squared modulus of the wavefunction, $|\psi_i(z)|^2$ for the first seven solutions of the \ch{LaAlO3}/\ch{SrTiO3}/\ch{LaAlO3} building block as a function of the \ch{SrTiO3} thickness. (a) and (b) represent thicknesses of 10.5 u.c. and 6.5 u.c. for the \ch{SrTiO3} layer, respectively. Light blue and green indicate the $d_{xy}$ (light) and $d_{xz}$ (heavy) bands. The Fermi level is indicated by a gray dashed line and the potential is shown as a dark blue line. The wavefunctions $|\psi_i|^2$ are shifted by their energies $E_i$. Note the different energy scales for (a) and (b). The solutions in the bottom of the well are twofold degenerate, with the asymmetric wavefunction higher in energy than the symmetric solution. The $z$ axis is the depth of the quantum well, along the stacking direction, with $0$ being the centre of the quantum well. \label{fig:schrodinger-possion}}
\end{figure}

The two 2DEGs with opposite Rashba splitting, within one building block, are assumed to be coupled by an \textit{intra}layer coupling,
\begin{equation}
D( \kv ) = (D_0 + Mk^2)I_{2\times 2},
\end{equation}
where $D_0$ is a constant, and a $Mk^2$ coupling term is introduced, as was suggested by Das and Balatsky \cite{dasEngineeringThreedimensionalTopological2013}. We will discuss this term in more detail in Sec.~\ref{sec:quadratic}, including how this coupling could be engineered. $I_{2\times2}$ accounts for spin. Subsequent building blocks are coupled with an \textit{inter}layer coupling, $t_z$. Both coupling terms are indicated in Fig.~\ref{fig:termination}(a). 

The Hamiltonian is then given by
\begin{equation}
H = \begin{pmatrix}
h_R^+ & D & 0 & 0 & \ldots\\
D & h_R^- & T & 0 & \ldots\\
0 & T & h_R^+ & D & \ldots\\
0 & 0 & D & h_R^- & \ldots\\
\vdots & \vdots & \vdots & \vdots & \ddots
\end{pmatrix},
\label{eq:ham-n-bilayers}
\end{equation}
where $T = t_z I_{2\times 2}$. 

\section{A single Rashba layer within the $\ch{LaAlO3}/\ch{SrTiO3}/\ch{LaAlO3}$ heterostructure}
\label{sec:lao-sto-lao-poisson}
Within a single building block it is possible to design a 2D topological insulator, similar to how the 2D topological insulator in a \ch{HgTe} quantum well \cite{bernevigQuantumSpinHall2006} is described by two coupled Dirac Hamiltonians \cite{buttnerSingleValleyDirac2011, tkachovTopologicalInsulatorsPhysics2015}.

\begin{figure*}
	\centering
	\includegraphics[width=15cm]{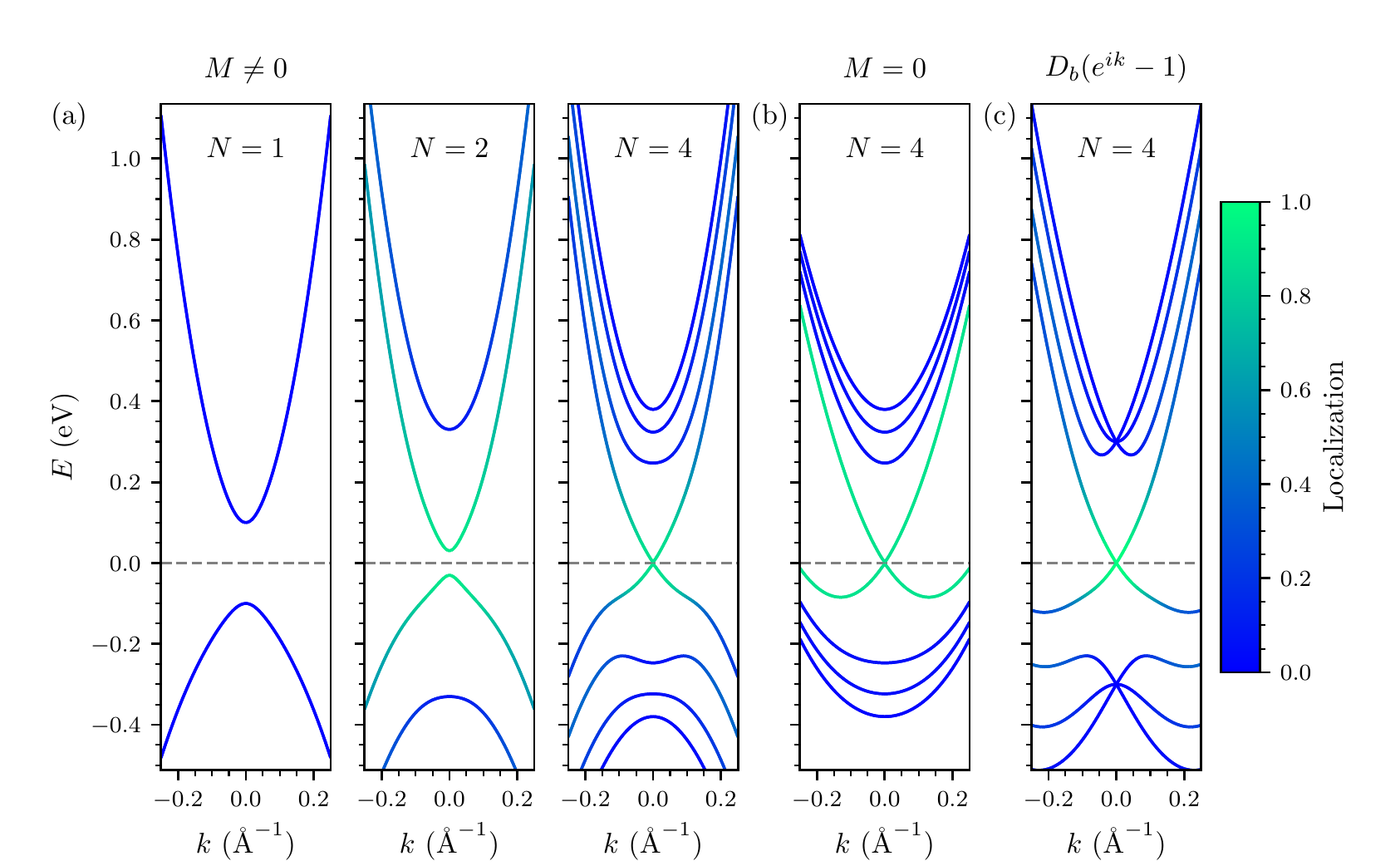}
	\caption{(a) The dispersion for the model given in Eq.~(\ref{eq:ham-n-bilayers}) for $N=1,2,4$ building blocks. The colorscale indicates the localization of the wavefunctions, defined as the modules squared $|\psi|^2$ on the top or bottom 2DEG. The dashed line shows that the surface states and bulk states are clearly separated. $m^*=\SI{0.1}{eV^{-1}\angstrom^{-2}}, D_0=\SI{-0.1}{eV}, t_z=\SI{-0.3}{eV}, \alpha_R=1.3, M=\SI{-10}{eV\angstrom^2}$  (b) Dispersion for $N=4$ layers, without the quadratic coupling $Mk^2$. Note that now the surface states are not separated from the bulk states, which means that both contribute to transport. (c) Dispersion for $N=4$ layers, with the coupling $D_b=\SI{-2}{eV}$. The coupling between the 2DEGs with opposite Rashba splitting is quadratic. Close to $k=0$ the bands are now well separated, but further away (not shown) the bulk bands curve upwards again.  $N$ is defined as the number of \ch{LaAlO3}/\ch{SrTiO3}/\ch{LaAlO3} building blocks.
		\label{fig:dispersion}}
\end{figure*}

Here, we solely consider a single \ch{LaAlO3}/\ch{SrTiO3}/\ch{LaAlO3} building block, described by the Hamiltonian
\begin{equation}
H_1=\begin{pmatrix}
h_R^+ & D \\
D & h_R^-
\end{pmatrix},
\end{equation}
which gives the dispersion as shown in Fig.~\ref{fig:dispersion}(a). 

This system has time-reversal symmetry with $\mathcal{T}^2$~=~$-1$ and inversion symmetry, so we apply the Fu-Kane criterion \cite{fuTopologicalInsulatorsInversion2007} and calculate the $\mathbb{Z}_2$ invariant \cite{tkachovTopologicalInsulatorsPhysics2015}. We assume the edges of the Brillouin zone, where three of the four time-reversal invariant momenta are located, to be trivial and only evaluate the parity at $k_x=k_y=0$ \cite{konigQuantumSpinHall2008b}. This is justified by the fact that within the \ch{LaAlO3}/\ch{SrTiO3} heterostructure the 2DEG is only present close to $\Gamma$ \cite{gariglioElectronConfinementLaAlO32015}. We find $(-1)^\nu = -1$ if $D(k=0) < 0$ and $(-1)^\nu =1$ if $D(k=0) > 0$, where $\nu$ is the $\mathbb{Z}_2$ invariant. Note that $D(k=0)=D_0$ and therefore the system is a 2D topological insulator if the coupling $D_0$ is negative. These edge states live in the gap with size $2|D_0|$. Note that the coupling $Mk^2$ which we introduced before does not play a role in determining the gap and the topological invariant, since it is only non-zero for $k>0$.

In order to explore a more explicit implementation of this system, we now focus on the formation of two 2DEGs within a single \ch{LaAlO3}/\ch{SrTiO3}/\ch{LaAlO3} heterostructure as depicted in Fig.~\ref{fig:termination}(b). Note that the method described here is general and can be applied to any interface where a 2DEG is present due to polar gating. Our goal is to determine how and when two 2DEGs can coexist on both interface of a \ch{SrTiO3} layer and the coupling between the two 2DEGs plays a role. In the limit of a thick \ch{SrTiO3} film interfaced with \ch{LaAlO3} at the top and bottom, we would expect a 2DEG to form at both interfaces. To be able to determine if these 2DEGs can coexist as the thickness of the \ch{SrTiO3} layer is reduced, we have to take both quantum mechanics and electrostatics into account. We do this by self-consistently solving the set of Schr\"odinger-Poisson equations \cite{gariglioElectronConfinementLaAlO32015,sminkGateTunableBandStructure2017b}.

We show the results of the first seven wavefunctions for the self-consistently solved system for varying thicknesses of the \ch{SrTiO3} layer in Fig.~\ref{fig:schrodinger-possion}. By tuning the thickness to 10.5~unit cell (u.c.), see Fig.~\ref{fig:schrodinger-possion}(a), we find that there are two coexistent 2DEGs, which couple to each other. The coupling manifests itself as a splitting of the wavefunction in a symmetric and asymmetric solution, where the energy difference between the two is the intralayer coupling strength. For 10.5 u.c., we found the strength of the coupling to be of the order $D_0~\simeq$~$-$3~meV, where we have determined by evaluating whether the symmetric or asymmetric solutions lie higher in energy.

For a system with a thinner \ch{SrTiO3} layer, see Fig. \ref{fig:schrodinger-possion}(b) for a thickness of 6.5 u.c., the potential well is no longer deep enough to localize the 2DEGs and we are essentially seeing the solution for a simple quantum well --- slightly perturbed by a small potential. In this case, the two 2DEGs can no longer coexist.

We note that changes in the physical parameters that enter our calculation, can affect the coexistence of the two 2DEGs. The most notable parameter is the permittivity of \ch{SrTiO3}, which is known to depend on the growth method \cite{christenDielectricPropertiesSputtered1994}. We found that a lower permittivity localizes the 2DEGs more towards the interface, and therefore can be used as a tuning parameter.

It is thus clear that in the \ch{LaAlO3}/\ch{SrTiO3}/\ch{LaAlO3} building block two 2DEGs can coexist and couple to each other, though weakly. It is the intralayer coupling which subsequently allows the system to become a 2D topological insulator --- as long as $D_0$ is negative. Moreover, by stacking this building block we can build a 3D topological crystalline insulator, as we will show next.

\section{Multilayer heterostructure}

We stack the \ch{LaAlO3}/\ch{SrTiO3}/\ch{LaAlO3} building blocks as described as given by the Hamiltonian in Eq.~(\ref{eq:ham-n-bilayers}) to create a multilayer heterostructure. In Fig.~\ref{fig:dispersion}(a) the dispersion is shown for one, two and four building blocks. The colorscale indicates localization on the top or bottom interface, defined as  the modulus squared of the wavefunction on the top or bottom 2DEG. For a single layer ($N$~=~1) there is clearly a gap, which closes as more layers are added. The closing of the gap is of course reminiscent of a topological insulator.

If we look at the localization of the eigenfunctions corresponding to the linear Dirac cone, we find that these are surface states. They are localized at either the top or bottom 2DEG and decay exponentially into the bulk. Most importantly, from the expectation value of the spin we find that these states are spin-momentum locked, such that for every energy there are two states: one with spin up and one with spin down, localized on the top and bottom of the structure. This system therefore has the hallmarks of a conventional 3D topological insulator, such as \ch{Bi2Te3}. 

The gap follows $\lim_{N\rightarrow\infty}\Delta= 2(|D_0| - |t_z|)$. Therefore,  the goal would be to engineer a strong enough difference between the intra- and interlayer coupling strengths. This can easily be achieved by increasing the thickness of the \ch{SrTiO3} layer, to control the distance between the opposite surfaces and thus the strength of the intralayer coupling. This is illustrated in Fig.~\ref{fig:termination}(a), where the \ch{SrTiO3} is much thicker than the \ch{LaAlO3} and thus $|D_0|$~$\ll$~$|t_z|$.

In order to classify the formation of the surface states, we introduce periodic boundary conditions along the $z$-direction by making the replacement $t^{\pm}_z\rightarrow t_z e^{\pm ik_z}$, where $+$ $(-)$ is the coupling to the layer above (below).
We classify the new system by looking at the symmetries present. We find time-reversal symmetry with $\mathcal{T}^2$~=~$-1$ and around $k$~=~0 a chiral symmetry $\mathcal{C}^2$~=~$+1$, which is broken as we move away from $k$~=~0. The combination of these two symmetries also implies particle-hole symmetry with $\mathcal{P}^2$~=~$+1$. We also find an additional reflection crystalline symmetry, given by the operator
\begin{equation}
	R_{\parallel}H\left( -k_x, -k_y, k_z  \right) R_{\parallel}^{-1} = H(k_x, k_y, k_z),
\end{equation}
with properties ${R}_{\parallel}^2=+1$ and $\left[ {R}_{\parallel},{\mathcal{C}} \right]=0$, where we define a reflection symmetry as any symmetry that involves $r\rightarrow \tilde{r}$ as a reflection symmetry, in accordance with the convention used by Chiu et al. \cite{chiuClassificationTopologicalQuantum2016}.

Then we can treat this system as effectively having two dimensions: the $z$-direction and the $x$-$y$ plane, which is reflection symmetric. A topological crystalline insulator in two dimensions in class AIII, with a reflection operator which commutes with the chiral symmetry, is classified by the mirror winding number $M\mathbb{Z}$ \cite{chiuClassificationTopologicalQuantum2016}. The mirror winding numbers are defined on the hyperplanes in the Brillioun zone which are symmetric under reflection. In our case the hyperplane is defined by $k_x$~=~$k_y$=~0. Block- diagonalizing the Hamiltonian at this hyperplane with respect to the two eigenspaces of the reflection operator ${R}_{\parallel}$ gives 
\begin{equation}
	AH(k_x=0, k_y=0,k_z)A^{-1} =   
	\begin{pmatrix}
	\bm{D}(k_z) & 0 \\ 
	0 & \bm{D}(k_z)
	\end{pmatrix},
\end{equation}
where
\begin{equation}
	\bm{D}(k_z) = \begin{pmatrix}
	0 & D_0 + t_ze^{-ik_z} \\
	D_0 + t_ze^{ik_z} & 0
	\end{pmatrix}
\end{equation}
and $A$ is the space formed by the eigenvectors of the reflection operator $R_{\parallel}$. This system is comparable to the Su-Schrieffer-Heeger (SSH) chain \cite{suSolitonsPolyacetylene1979}. We apply a similar analysis, finding a winding number
\begin{equation}
	Q(H) = \begin{cases}
	1\quad \text{if } |D_0| < |t_z|, \\
	0\quad \text{if } |D_0| > |t_z|.
	\end{cases}
\end{equation}

The multilayer heterostructure is a 3D topological crystalline insulator if $Q(H)=1$, so if the intralayer coupling, $D_0$, is weaker than the interlayer coupling, $t_z$. In other words: the surfaces with opposite Rashba splitting 2DEGs need to placed relatively far apart, such that the intralayer coupling is weak, and then stacked on top of each other, which induces a strong interlayer coupling. This is illustrated in Fig.~\ref{fig:termination}(a).

Here, we would like to stress that this is not a 3D topological insulator protected by time-reversal symmetry, such as \ch{Bi2Te3}, even though the system could also be classified in class AII. Instead, this is a 3D topological crystalline insulator, where the topology is protected by a combination of chiral symmetry and reflection symmetry. This is in contrast to the 2D topological insulator we found within a single Rashba layer, as described in Sec.~\ref{sec:rashba-bilayers}, which is a conventional class AII topological insulator, protected by time-reversal symmetry.

Moreover, this multilayer heterostructure can still be a 2D topological insulator if $Q(H)$~=~0 with the $\mathbb{Z}_2$ invariant given by the Fu-Kane criterion \cite{fuTopologicalInsulatorsInversion2007}, analogous to the analysis of the single \ch{LaAlO3}/\ch{SrTiO3}/\ch{LaAlO3} building block system as was done in Sec.~\ref{sec:lao-sto-lao-poisson}. Applying this invariant, we found that the multilayer system oscillates between a trivial and a 2D topological insulator as a function of the number of layers if $|D_0|>|t_z|$, i.e. the system is gapped everywhere and surface states do not form. For one building block the system is a 2D topological insulator, as was shown before. For two and three building blocks ($N$~=~2,~3) the system is a trivial insulator and for four and five building blocks ($N$~=~4,~5) the system again becomes a 2D topological insulator, and so on, switching every second added bilayer.

\begin{figure*}
	\centering
	\includegraphics[width=15cm]{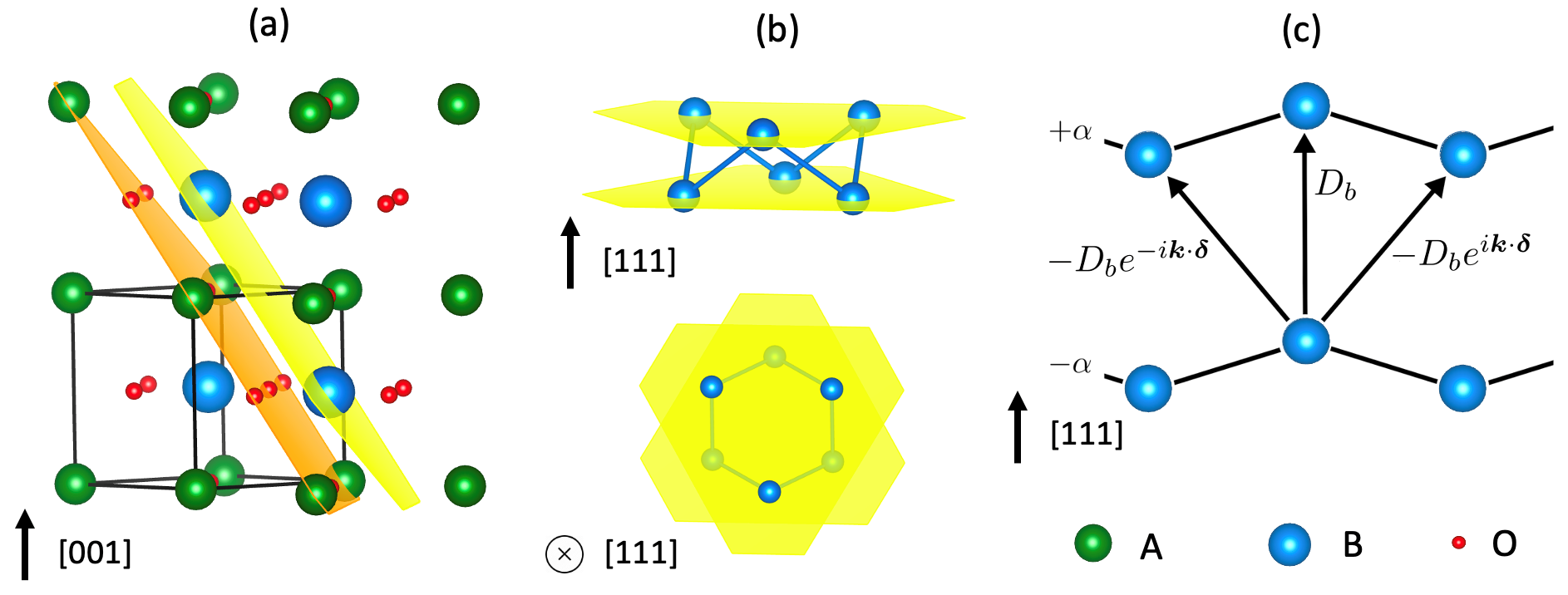}
	\caption{Schematic illustration how a buckled honeycomb forms in a [111]-oriented \ch{ABO3} perovskite structure. The green, blue and red dots represent the A, B and O atoms, respectively. The sizes are not realistic atom radii. (a) Shows the \{111\} planes intersecting with the A- and B-cations, respectively for the orange and yellow plane in a perovskite structure with the [001] direction along the vertical axis. (b) Shows how the buckled honeycomb structure forms when two [111]-oriented layers are stacked on top of each other. In the top image, the B atoms are shown with the [111] direction along the vertical axis. In the bottom image, the [111] direction is pointing in the plane, away from the viewer. Here, the rotation of the two layers with respect to each other is visible. A clear buckled hexagonal structure is observed. In (c) a possible implementation of the quadratic coupling is shown. The second neighbour coupling picks up an additional phase. For simplicity we assume that the strength of the coupling for the nearest neighbour and second nearest neighbour is the same.
		\label{fig:BBO111}}
\end{figure*}

\section{Separating the surface and bulk bands}
\label{sec:quadratic}
It is necessary to analyse, in more detail, the Hamiltonian as introduced by \textcite{dasEngineeringThreedimensionalTopological2013} and most importantly the quadratic coupling, $Mk^2$, which couples the 2DEGs with opposite Rashba splitting. Naively, one would expect a simple, constant coupling. A notable exception is the coupling between the two 2DEGs formed on the top and bottom surfaces of the topological insulators \ch{Bi2Te3}, \ch{Bi2Se3} and \ch{SbTe3} when studied as thin films \cite{asmarInterfaceSymmetrySpin2017,asmarTopologicalPhasesTopologicalinsulator2018}. As the thickness of the film is reduced, the coupling starts to oscillate, which is  responsible for the supposed switching between a 3D and 2D topological insulator \cite{liuOscillatoryCrossoverTwodimensional2010}. Note that this is not related to the switching we observed when stacking multiple building blocks. 

Considering a heterostructure of four \ch{LaAlO3}/\ch{SrTiO3}/\ch{LaAlO3} building blocks ($N$~=~4) without the quadratic coupling results in the dispersion as presented in Fig.~\ref{fig:dispersion}(b), where around $k=0$ the Dirac cone is still present. However, the bands curve up away from $k=0$, which turns the system into a bulk conductor. In transport experiments, the contribution from the bulk bands would completely drown out the surface states, making them inaccessible. Thus, without the $Mk^2$ coupling term between the 2DEGs with opposite Rashba splitting, this system is no longer a topological (crystalline) insulator, neither in two or three dimensions.

The quadratic coupling, therefore, serves as a useful method to separate the surface and bulk bands in our system, but there are more factors at play. Firstly, it is important to note that the curvature of the bands, which is controlled by the effective mass, $m^*$, plays an important role. Shallower bands will curve up less before reaching the end of the Brillouin zone, and thus allowing for a gap throughout the complete Brillouin zone. Within the \ch{LaAlO3}/\ch{SrTiO3}/\ch{LaAlO3} heterostructure it is known that strain can be used to tailor the 2DEGs properties \cite{barkTailoringTwodimensionalElectron2011}. More recently, it was also shown that it is possible to synthesize free-standing \ch{SrTiO3} thin films \cite{luSynthesisFreestandingSinglecrystal2016}, with which it would be possible to further investigate the effect of strain on the 2DEG properties.

Secondly, we found numerically that the Rashba coupling also serves to separate the bands. However, the Rashba coupling is much more difficult to engineer, since it is essentially an intrinsic quality of the materials used through the strength of the spin-orbit coupling and the interface potential. Note that in the systems we propose, the electric potential is opposite at the interfaces, so an external potential cannot be used to increase the interfacial potential, as has been suggested for some polar interfaces \cite{shanavasElectricFieldTuning2014}.
As already stated, we are not necessarily bound to the \ch{LaAlO3}/\ch{SrTiO3} system, but our results are more general. Other interfaces, such as \ch{LaVO3}/\ch{KTaO3} \cite{wadehraPlanarHallEffect2020}, could possibly have flatter bands and higher Rashba splitting, which would realize the appropriate band separation. An alternative approach, as used by Trifunovic et al. \cite{trifunovicCoupledRashbaElectron2016a}, is to make use of alternating electron and hole 2DEGs, which naturally have bands with opposite curvatures.

We return to the quadratic coupling, as introduced by Das ans Balatsky \cite{dasEngineeringThreedimensionalTopological2013}. The most straightforward way engineer this coupling term, is by making use of the geometry of the system. We propose to use a coupling of the form $D(k)$~$=$~$D_b\left( e^{ik} -1 \right)$ \cite{fulgaCoupledlayerDescriptionTopological2016}. From a Taylor expansion, it is easy to see that this kind of coupling goes as $D(k)$~$\propto$~$ik-k^2/2+\mathcal{O}(k^3)$. We show the resulting dispersion in Fig.~\ref{fig:dispersion}(c), where we can see that the bands are well separated. 

As is clear from the Taylor expansion and the dispersion, this coupling is only quadratic around $k=0$. This type of coupling will only serve to separate the bands in some region of the Brillioun zone, and not throughout. The bands as shown in Fig.~\ref{fig:dispersion}(c), therefore, curve upwards for larger $k$, which is not shown here. More work is needed to understand the behaviour throughout the complete Brillouin zone.

We speculate that achieving this kind of coupling can be done by introducing a second neighbour coupling, as depicted in Fig.~\ref{fig:BBO111}(c). Here, two buckled 2DEGs are coupled in such a way that a unit cell not only couples to the ones directly above it, but also to unit cells adjacent to it. The coupling has the form $\exp\left[i\kv \cdot \bm{\delta}\right]$, where $\bm{\delta}$ is the vector connecting two unit cells. This does, however, require a switch in the sign between nearest and next nearest neighbour coupling. One possibility to achieve this, is if one coupling is formed by a $\sigma$ bond and the other coupling by a $\pi$ bond. For certain materials it is known that these bonds have couplings with opposite sign \cite{katkovHoppingParametersTunnel2018}.

The coupling of the form $D(k)=D_b\left( e^{ik} -1 \right)$ has been predicted to exist in multiple systems, most notably in coupled buckled honeycomb lattices \cite{kooiInversionsymmetryProtectedChiral2018, mccannLandauLevelDegeneracyQuantum2006,xiaoInterfaceEngineeringQuantum2011}.
These systems include the buckled honeycomb formed by the [111]-bilayer, which has been shown to exhibit the required Rashba-type splitting \cite{kimGrapheneAnalogue1112018}. We are not aware of the existence of this kind of coupling in the often used \ch{LaAlO3}/\ch{SrTiO3} (001) heterostructure, so next we will elaborate further on how the [111]-bilayers might offer possible implementations of the model described in this work.

\section{The [111]-bilayer examined}
\label{sec:outlook}
In [111]-bilayers, topological phases are established by making smart use of new design principles. Most of the theoretical modelling where electron-electron correlations and topology are combined have been performed for honeycomb lattices. The original Haldane model \cite{haldaneModelQuantumHall1988}, in order to realize a Chern insulator, was also developed for a honeycomb lattice. It turned out that the spin-orbit coupling is too small to open a gap of substantial size in graphene. The use of heavier transition metal or rare earth metal ions is interesting in this respect. 

A two-unit cell thick perovskite film with the structural formula \ch{ABO3}, stacked along the [111] direction, forms a buckled honeycomb structure. This is schematically depicted in Fig.~\ref{fig:BBO111}. The orange and yellow planes in Fig.~\ref{fig:BBO111}(a) show the A- and B-cations that lie in the \{111\} plane, respectively, with respect to the unit cell with the [001] direction along the vertical axis. Both A- and B-site cations form a triangular lattice in a single \{111\} plane. However, as illustrated, the orange plane also intersects with the oxygen atoms, where the yellow plane solely contains B-site cation atoms.

When two perovskite unit cells are stacked in the [111] direction, the B-site cations form a buckled honeycomb structure as schematically shown in Fig.~\ref{fig:BBO111}(b). Where the nearest neighbour B-cations are now connected by a blue line. In the upper image, the [111] direction is along the vertical axis, so that the structure is depicted as a side view. It is clear that the structure is buckled, implying a spacial separation between two nearest B-site cations. The lower image shows the B-sites from the top, where the rotation with respect to each other becomes visible. Experimentally, a [111]-oriented bilayer can be realized by sandwiching two unit cells between another perovskite structure. By using the same atom as in the bilayer for the A-site, the two B-site \{111\} planes in the bilayer are isolated from the rest of the structure, creating the buckled honeycomb.

These [111]-bilayer perovskites form a rich playground for investigating theoretical predictions \cite{wrightRealisingHaldane2013}. From theoretical calculations based on tight binding \cite{xiaoInterfaceEngineeringQuantum2011} and DFT + U models \cite{doennigDesignOfChern2016, wengTopologicalMagnetic2015, guoWideGapChern2017} (i.e. explicitly taking correlations into account), QAHE topological phases have been predicted for heterostructures where a bilayer of \ch{LaBO3} is embedded in \ch{LaB'O3}, where B and B' are transition metals.  Ion B' (such as Al or Ni) determines the lattice constant by epitaxial strain, and ion B determines the topology. Most promising combinations with B'~=~Al are B~=~Mn, Os and Ru \cite{doennigDesignOfChern2016, wengTopologicalMagnetic2015, guoWideGapChern2017}.  

For B'~=~Mn, \ch{LaMnO3} has the disadvantage that a possible Jahn-Teller distortion would break the symmetry of the lattice so that the topological phase is lost. Experiments with different levels of strain, to possibly overcome the Jahn-Teller distortion, will have to tell if the topological insulating phase can still be established. The heavier 4d and 5d transition metals will not show this symmetry breaking so strongly. The choice for the 5d ion B~=~Os will perhaps be difficult to stabilize, but B~=~Ru might be realistic. Given the challenges outlined in the previous sections, we deem the experimental realization of [111]-oriented bilayers very promising.

\section{Conclusion}
In conclusion, 2D and 3D topological states can be realized when stacking the building block \ch{LaAlO3}/\ch{SrTiO3}/\ch{LaAlO3} in which a 2DEG is present on the top and bottom interface, having opposite Rashba splitting. The two 2DEGs are coupled via a quadratic coupling term, $Mk^2$. When considering a single building block, the system is a 2D topological insulator when the coupling, $D(k=0)$, is negative. When stacking more than one building block, both the intra- and interlayer coupling play an important role. When the intralayer coupling,  $D_0$, is smaller than the interlayer coupling, $t_z$, the system is a 3D topological crystalline insulator. When its the other way around, the system is alternately a trivial and 2D topological insulator, depending on the number of building blocks. 

To isolate the surface from the bulk states, it is beneficial to have shallower bands and higher Rashba splitting. Also, the quadratic coupling term, as introduced, creates a band separation. Experimentally, this can be achieved in a buckled honeycomb lattice, for example, a [111]-oriented bilayer in a perovskite structure.

The two approaches towards achieving topological oxides described in this work, 1) stacking oxide Rashba layers and 2) using [111]-bilayer perovskite oxides, have become very much related. Although no experimental results of topological nature have been achieved in either of the two research directions, we believe that all technological ingredients are in place to verify the theoretical predictions in the near future. An important asset in the field of oxide heterostructures is the sub-unit cell level of control over the required materials. And, finally, the class of oxide materials provides a large variety of available parameters, such as buckling, strain and different orbitals to tune the electronic bands of a heterostructure into the topological regime. 

\section*{Acknowledgements}
This work was financially supported by the Netherlands Organization for Scientific Research (NWO) through a VICI grant (680-47-625) and an ENW-Groot grant (Topcore). The numerical data that support the findings of this study are available from the corresponding author upon reasonable request.

%

\end{document}